\RequirePackage[2020-02-02]{latexrelease}
\RequirePackage{fix-cm}
\documentclass{svjour3}                     %
\def\makeheadbox{\relax}

\usepackage{cite}
\usepackage{amsmath,amssymb,amsfonts}
\usepackage{algorithmic}
\usepackage{graphicx}
\usepackage{textcomp}
\usepackage{xcolor}
\usepackage[hyphens]{url}
\usepackage{hyperref}
\usepackage{listings}
\usepackage{booktabs}
\usepackage{makecell}
\usepackage{enumitem}
\usepackage{multirow}
\usepackage{tabularx}
\usepackage{newfloat}
\usepackage[acronym,smallcaps]{glossaries}
\usepackage{bbding}
\usepackage[framemethod=TikZ]{mdframed}
\usepackage[misc]{ifsym}
\mdfdefinestyle{OBS}{%
    leftline=false,
    rightline=false,
    skipabove=4pt,
    frametitle={Observations},
    frametitleaboveskip=4pt,
    frametitlebelowskip=2pt,
    innertopmargin=2pt,
    innerbottommargin=2pt,
    innerrightmargin=0pt,
    innerleftmargin=0pt
}

\newcommand{\observations}[1]{

  \vspace{0.2em}
  \hrule 
  \vspace{0.2em}

  \noindent

  {\bf \small Observations. }#1
  \vspace{0.2em}
  \hrule
  \vspace{0.2em}
}

\newcommand\circled[1]{\raisebox{.5pt}{\textcircled{\raisebox{-.9pt} {#1}}}}

\newacronym{td}{td}{technical debt}
\newacronym{vs}{vs}{Visual Studio}
\newacronym{sq}{sq}{SonarQube}
\newacronym{mi}{mi}{Maintainability Index}
\newacronym{hv}{hv}{Halstead Volume}
\newacronym{cc}{cc}{Cyclomatic Complexity}
\newacronym{loc}{LOC}{Lines of Code}
\newacronym{sig}{sig model}{SIG/TÜViT Evaluation Criteria Trusted Product Maintainability}
\newacronym{bch}{BCH}{Better Code Hub}
\newacronym{aip}{AIP}{CAST Application Intelligence Platform}
\newacronym{ccq}{ccq}{Code Climate Quality}
\newacronym{squore}{SSA}{SQUORE Software Analytics}
\newacronym{ci}{CoIn}{Code Inspector}
\newacronym{cs}{CoSc}{CodeScene}
\newacronym{vcs}{VCS}{version control system}
\newacronym{kiuwan}{KQA}{Kiuwan Code Analysis (QA)}
\newacronym{sig_name}{SIG}{Software Improvement Group}
\newacronym{ui}{UI}{user interface}

\def\smath#1{\text{\scalebox{.8}{$#1$}}}

\def\sfrac#1#2{\smath{\frac{#1}{#2}}}

\begin{document}
\title{Technical Debt and Maintainability: \\How do tools measure it?}

\author{Rolf-Helge Pfeiffer         \and
        Mircea Lungu %
}

\institute{R.-H. Pfeiffer (\Letter) \at
              IT University of Copenhagen
              Rued Langgaards Vej 7 \\
              Tel.: +45-7218 5102\\
              \email{ropf@itu.dk}
           \and
           M. Lungu \at
              IT University of Copenhagen
              Rued Langgaards Vej 7 \\
              \email{mlun@itu.dk}
}

\date{Version 1.0: Dec. 2019, Version 1.1: Aug. 2020}

\maketitle

\begin{abstract}
The technical state of software, i.e., its \textit{\gls{td}} and \emph{maintainability} are of increasing interest as ever more software is developed and deployed. Since \textit{\gls{td}} and \emph{maintainability} are neither uniformly defined, not easy to understand, nor directly measurable, practitioners are likely to apply readily available tools to assess \gls{td} or \emph{maintainability} and they may rely on the reported results without properly understanding what they embody. In this paper, we: \textit{a)} methodically identify 11 readily available tools that measure \gls{td} or \emph{maintainability}, \textit{b)} present an in-depth investigation on how each of these tools measures and computes \gls{td} or maintainability, and \textit{c)} compare these tools and their characteristics. We find that contemporary tools focus mainly on internal qualities of software, i.e., quality of source code, that they define and measure \gls{td} or maintainability in widely different ways, that most of the tools measure \gls{td} or maintainability opaquely, and that it is not obvious why the measure of one tool is more trustworthy or representative than the one of another.
\end{abstract}

\section{Introduction}\label{sec:intro}

The Danish Agency for Digitization, under the Danish Ministry of Finance, prescribes a model for portfolio-management of central government IT systems~\cite{digst2019model}, which requires all public agencies to map the ``technical state'' of central government IT systems for triennial review. One of the questions in the model asks the agencies to rate \emph{``How satisfactory is the technical state of the IT system today?''} for all respective IT systems under their responsibility. The question must be answered on a scale ranging from 1 (very unsatisfactory) to 5 (extremely satisfactory). The manual for the model explains that the answer should be given by assessing each system's \gls{td} and the \emph{maintainability} (\emph{``possibility to further develop and integrate the system''}). However, the manual does neither specify how to assess \gls{td} or maintainability of a system nor what precisely is meant by these terms. Since neither of the two concepts is easy to understand nor directly measurable, managers and software developers are likely to rely on readily available tools to assess \gls{td} or maintainability. Some Danish agencies, such as the tax agency or the court administration, have used commercially available tools, for example those provided by the \gls{sig_name}\footnote{\url{https://www.softwareimprovementgroup.com}} to assess \gls{td} and maintainability of software systems~\cite{ft2015efi,ft2013jfs}.

However, there exists a multitude of tools that are reported to be suitable for identification and management of \gls{td} or maintainability, see \autoref{sec:study}. In practice, questions arise, such as, ``How to understand the values or ratings that various tools produce for \gls{td}/maintainability?'', or ``How are these values and ratings actually measured and computed?'', i.e., ``What do they represent?''. We translate these generic questions into research questions for this paper:

\begin{description}
  \item[RQ1] How do tools define the concepts \gls{td} or maintainability?
  \item[RQ2] How do tools measure and compute values for \gls{td} or maintainability?  
\end{description}

To study our research questions, we identify a comprehensive list of 11 readily available tools, see~\autoref{sec:study}, of which six are not covered in previous academic studies~\cite{rios2018tertiary, fontana2016technical,lenarduzzi2018survey}. Unlike previously studied tools, those included in this paper actually assess \gls{td} or maintainability. That is, they report direct measurements and ratings for these two concepts. For each tool, we provide a precise description of how these two concepts are defined, measured, and computed.

After investigating RQ1 and RQ2 (\autoref{sec:results}) we report (\autoref{sec:discussion}) that: 
    \emph{a)} not all tools define explicitly \gls{td} or maintainability even though they report measurements for these concepts, 
    \emph{b)} tools define and measure \gls{td} or maintainability in widely different ways, 
    \emph{c)} tools focus mainly on internal quality, i.e., quality of source code, and \emph{e)} most of the tools measure \gls{td} or maintainability opaquely, so that it is not directly accessible why the measure of one tool is more representative than the one of another.

Note, other organizations than the Danish Agency for Digitization are interested in assessing \gls{td} or maintainability of software too. We believe that a nation-wide institutionalized requirement to periodically assess \gls{td} or maintainability of software provides sufficient motivation to gain a thorough understanding of the above research questions. Thus, the goal of this paper is:

\begin{description}
 \item[a)] to allow practitioners to better understand if and to which degree the reports of certain tools are suitable for assessing \gls{td} and maintainability in their domain, and 
 \item[b)] to collect information that is otherwise scattered around a plethora of sources to gain a quick overview of how certain tools measure and compute \gls{td} and maintainability.
\end{description}

The goal of this paper is \textbf{not} to \textit{systematically} identify as many tools as possible that assess \gls{td} and maintainability and the goal is \textbf{not} to \textit{recommend} which tool to use to assess \gls{td} and maintainability.

\section{Background \& Related Work}\label{sec:background}

The \emph{\gls{td}} concept was coined in 1992~\cite{cunningham1992wycash}, as: \textit{``Shipping first time code is like going into debt. A little debt speeds development so long as it is paid back promptly with a rewrite. [\ldots] The danger occurs when the debt is not repaid. Every minute spent on not-quite-right code counts as interest on that debt.''}~\cite{cunningham1992wycash}. It is a metaphor and it is imprecise by nature as it builds upon the assumption that we know what \emph{quite-right} code is. 
However, we do not have generally true measures for software quality, i.e., \emph{quite-right} code~\cite{kitchenham1996software,boehm1976quantitative}. 
Often code quality is domain specific as for example stated in the ISO/IEC 25000 standard: \emph{``for interactive consumer software, such as word processor, usability and co-existence with other software [...] is considered important. For Internet and open systems, security and interoperability are most important.''}~\cite{iso25000}.

Likely, due to being a metaphor, the term \gls{td} was defined differently by various authors since its initial occurrence. Some authors define it solely in terms of source code~\cite{sappidi2010cast,curtis2012estimating,curtis2012estimating_ws}, e.g., as \emph{``cost of the effort required to fix problems that remain in the code when an application is released to operation.''}~\cite{sappidi2010cast}

Others, consider \gls{td} more broadly than to be just caused by properties of code~\cite{ernst2012role,avgeriou2016managing}, e.g., 
Avgeriu et. al. consider it as \emph{``[\ldots] collection of design or implementation constructs that are expedient in the short term, but set up a technical context that can make future changes more costly or impossible.''}~\cite{avgeriou2016managing}.
Some practitioners consider \gls{td} more pragmatically and even more general. For example, for Birchall \gls{td} is \textit{``a metaphor for the accumulation of unresolved issues in a software project''}~\cite{birchall2016re}. and for Radigan from Atlassian it is \emph{``the difference between what was promised and what was actually delivered.''}\footnote{\url{https://www.atlassian.com/agile/software-development/technical-debt}}. 

Often practitioners rely on tools to quantify \gls{td}, where it is unclear how precisely the tools measure and compute \gls{td}. Tornhill presents an anecdote to illustrate the ``perils'' of quantifying \gls{td}: \textit{``I visited an organization to help prioritize its technical debt. [\ldots] the team had evaluated a tool capable of quantifying technical debt. The tool [\ldots] estimated how much effort would be needed to bring the codebase to a perfect score [\ldots] and the tool reported that they had accumulated 4\,000 years of technical debt!''}~\cite{tornhill2018software} What do these 4\,000 years of \gls{td} mean? Where do they stem from and how are they computed? This is the motivation for this paper. We want to better understand how contemporary tools quantify \gls{td} or maintainability and create values like the afore-mentioned one. 

Similar in goal to this study, Fontana et al.~\cite{fontana2016technical} investigate how five tools define and measure \gls{td} with respect to conformance to architecture. The paper does not describe how the tools were selected, except of being known to the authors. Furthermore, some of the studied tools do not measure \gls{td} as such but perhaps related concepts, such as, \emph{Excessive Structural Complexity} (Structure101), \emph{Structural Debt} (Sonargraph), and a \emph{Quality Deficit Index} (inFusion). 

In this study, we rigorously identify tools that actually assess \gls{td} or maintainability.
A recent tertiary study~\cite{rios2018tertiary} identifies 31 tools for identification and measurement of \gls{td}. 
Since it is a tertiary study, the authors list all tools that are reported in secondary studies, leading to inclusion of concepts, such as, \emph{continuous integration} or visualization techniques such as \emph{TD Board}~\cite{dos2013visualizing} or \emph{Code Christmas Trees}~\cite{kaiser2011selling} without being clear in which way these precisely identify or measure \gls{td}. In this study, we consider all the listed tools again and identify only those that actually measure \gls{td} or maintainability.

Similarly, Lenarduzzi et al.~\cite{lenarduzzi2018survey} identify 25 tools that support software maintenance tasks. The authors categorize each tool into at least on main goal category, such as, bug detection, testing, code review, etc. However, from the study it is not clear to which degree the mentioned tools are suitable for assessing maintainability or to which degree they are just meant to support the goal for which they were categorized.

Ernst et al.~\cite{ernst2015measure} study conceptions of \gls{td} amongst software professionals revealing that --~except architecture~-- engineers do not agree on the sources of \gls{td}, that less than 20\% consider problems in source code as origin of it, and that \gls{td} assessment tools are rarely used since interpreting results is too complex. With our study we hope to complement that work by making explicit what such tools consider \gls{td} or maintainability and how these are measured and aggregated.

\begin{figure*}
  \centering
  \includegraphics[width=\linewidth]{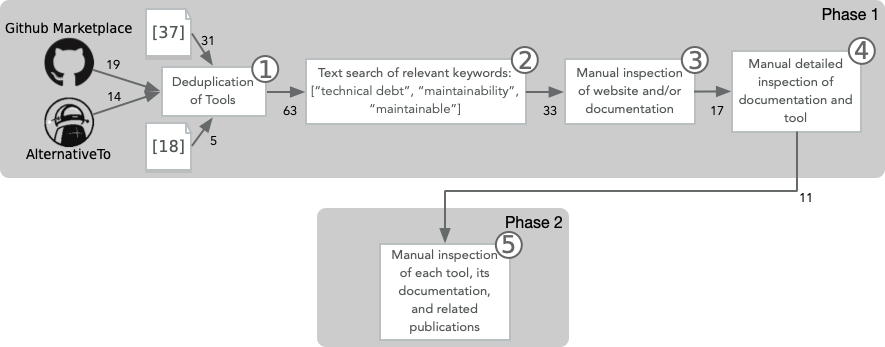}
  \caption{Study setup.}
  \label{fig:pipe}
\end{figure*}

\section{The Study}\label{sec:study}

In this study, we address the following research questions:

\begin{description}
\item[RQ1] How do tools define the concepts \gls{td} or maintainability?
\item[RQ2] How do tools measure and compute values for \gls{td} or maintainability?  
\end{description}

Our study is executed in two phases, see \autoref{fig:pipe}. Phase 1 identifies tools that actually assess \gls{td} or maintainability from literature and industrial sources. Phase 2 studies the identified tools in-depth to answer our research questions.

\subsection{Methodology}\label{subsec:method}

To identify tools for assessment of \gls{td} or maintainability we rely on four sources\footnote{Note, all links provided in this paper were last accessed in Nov. 2019}: 
\emph{a)} a recent tertiary study on \gls{td}~\cite{rios2018tertiary}, 
\emph{b)} a study by Fontana of five tools measuring aspects of \gls{td}~\cite{fontana2016technical}, 
\emph{c)} code quality tools from the GitHub Marketplace\footnote{\url{https://github.com/marketplace}}, and 
\emph{d)} \emph{AlternativeTo}, a crowd-sourced list of tools, which we queried for alternatives to ``SonarQube''\footnote{\url{https://alternativeto.net/software/sonarqube/}}. We select these sources for the following reasons: 
\begin{description}

    \item The Fontana et al. study~\cite{fontana2016technical} it is --~to the best of our knowledge~-- the only study on how tools measure and compute \gls{td} and related concepts.

    \item The tertiary study~\cite{rios2018tertiary} collects tools for identification and measurement of \gls{td} from 13 secondary studies (covering 703 primary sources) and such a comprehensive list is likely to include all potential tools. 

    \item The GitHub marketplace is one of the most popular collection of Software as a Service (SaaS) development tools.
    
    \item The crowd-sourced platform \emph{AlternativeTo}, is meant to find tools with comparable features. We know from our previous professional work that practitioners find relevant tools based on similarity recommendations. As a starting point for finding alternatives we use the widely popular quality assessment tool SonarQube~\cite{garcia2016improved}.
\end{description}

Based on the four sources, we follow a four staged process to identify tools that \emph{actually} measure and compute \gls{td} or maintainability, see \autoref{fig:pipe}. From Rios et al.~\cite{rios2018tertiary}, we collect all 31 tools that are listed under the categories \emph{identification} and \emph{measurement} of \gls{td} in appendix F. From Fontana et al.~\cite{fontana2016technical}, we collect all five studied tools. From the Github marketplace\footnote{\url{https://github.com/marketplace}}, we collect all 19 unique verified tools (which include commercial tools\footnote{\url{https://developer.github.com/marketplace/\#unverified-apps}}) from the categories \emph{code quality} and \emph{code review}.
From AlternativeTo, we collect 14 tools that are listed as similar to SonarQube.

After deduplication (\circled{1} in \autoref{fig:pipe}) of tools from the four sources the number of unique tools is 63. For each of these, we perform a text search via Google on the corresponding website or online documentation for the three search terms \emph{``technical debt''}, \emph{``maintainability''}, \emph{``maintainable''} (\circled{2} in \autoref{fig:pipe}). The Google query is of the form: {\footnotesize \texttt{site:<website> "<search term>"}}
Only tools for which we can find ``hits'' to at least one of the three search terms (only true for 33 tools) are kept for a manual inspection of the corresponding website or documentation page (\circled{3}). As many tools and vendors use \gls{td} and maintainability in blog posts, for marketing, etc., we exclude in stage \circled{3} all those tools for which we cannot find evidence, e.g., screenshots, formulas, etc., on the websites/documentation that they provide an \textbf{actual measure} of \gls{td} or maintainability, leaving 17 tools. By \textbf{actual measure}, we mean that the tools have to provide a measure that is literally labeled with \emph{``technical debt''} or \emph{``maintainability''}, i.e., it is not up to a users' interpretation if a certain result represents either of the two. In case we are in doubt after checking the online resources, we briefly experiment with the remaining 17 tools (\circled{4}) to see if an actual measure of \gls{td} or maintainability is provided in the tool's \gls{ui}. After which we are left with 11 tools that form the basis of our study in Phase 2. The 11 identified tools in alphabetical order are:

\begin{itemize}
    \item [A.] Better Code Hub {\footnotesize (\url{https://www.bettercodehub.com})}, 
    \item [B.] CAST AIP {\footnotesize (\url{https://www.castsoftware.com/products/application-intelligence-platform})}, 
    \item [C.] Code Climate Quality  {\footnotesize (\url{https://codeclimate.com})}, 
    \item [D.] Code Inspector  {\footnotesize (\url{https://www.code-inspector.com})}, 
    \item [E.] Codescene {\footnotesize (\url{https://codescene.io})}, 
    \item [F.] Kiuwan Code Analysis (QA) {\footnotesize (\url{https://www.kiuwan.com/code-analysis-qa})}, 
    \item [G.] NDepend {\footnotesize (\url{https://www.ndepend.com})}, 
    \item [H.] SonarQube  {\footnotesize (\url{https://www.sonarqube.org})}, 
    \item [I.] SQUORE Software Analytics {\footnotesize (\url{https://www.vector.com/int/en/products/products-a-z/software/squore/squore-software-analytics-for-project-monitoring})}, 
    \item [J.] SymfonyInsight  {\footnotesize (\url{https://insight.symfony.com})}, and 
    \item [K.] Visual Studio  {\footnotesize (\url{https://docs.microsoft.com/en-us/visualstudio/?view=vs-2019})}.
\end{itemize}

We study each of the 11 tools in depth based on the available documentation, related publications, and the tools themselves, i.e., we experiment with them\footnote{ Only for CAST AIP and Visual Studio we do not experiment with the tools themselves.}. The results of this step (\circled{5}) are presented per tool in the following section.

\section{Results}\label{sec:results}

In the following sections, we answer for each of the 11 tools the two research questions
\textit{RQ1} and \textit{RQ2}. Each section starts with how the respective tool \emph{defines} the concepts \gls{td} or maintainability after which, we describe how the respective tool \emph{measures} and \emph{computes} values for \gls{td} or maintainability.

\subsection{Better Code Hub}\label{subsec:bch}

\gls{bch} does neither define nor measure \gls{td}. Neither its documentation\footnote{\label{url_conf_man}\url{https://bettercodehub.com/docs/configuration-manual}, \url{https://bettercodehub.com/docs/faq}} nor corresponding publications~\cite{visser2016building,heitlager2007practical} mention the concept. But, \gls{bch} defines \emph{maintainability} as the property of \emph{``how easily a system can be modified''}~\cite{visser2016building} and measures it as a ratio of compliance to ten guidelines --~the \gls{sig}~\cite{sig10criteria,visser2016building,heitlager2007practical}. 

\gls{bch} measures the internal quality characteristic \emph{maintainability} and its five sub-characteristics of modularity, reusability, analyzability, modifiability, and testability according to ISO/IEC 25010~\cite{iso25010_2011} based on ten ``guidelines'' on 17 languages ranging from C\# over PHP to Kotlin\textsuperscript{\ref{url_conf_man}}. Each sub-characteristics is measured by a metric whose implementation is not available. The guidelines are: \emph{a) Write Short Units of Code}, \emph{b) Write Simple Units of Code}, \emph{c) Write Code Once}, \emph{d) Keep Unit Interfaces Small}, \emph{e) Separate Concerns in Modules}, \emph{f) Couple Architecture Components Loosely}, \emph{g) Keep Architecture Components Balanced}, \emph{h) Keep Your Codebase Small}, \emph{i) Automate Tests}, and \emph{j) Write Clean Code}. Code can either comply or not comply to any of these ten guidelines. 

The tool uses thresholds to decide if guidelines are respected or not. For example, to comply with \emph{a) Write Short Units of Code} the following distribution for LOC per unit (functions, methods, etc.) has to hold: at least 56.7\% of all units have $LOC\leq15$, at most 21.4\% of all units have $15<LOC\leq30$, at most 15.4\% of all units have $30<LOC\leq60$, and at most 6.9\% of all units have  $LOC>60$. We extracted these thresholds by inspecting the tool's UI during the analysis of a Java system (Apache Commons VFS) since no other resource specifies them explicitly. It is unclear if they are the same for other systems and other languages.

\gls{sig_name} offers another commercial version on \gls{bch} called \emph{Sigrid}~\cite{baggen2012standardized,heitlager2007practical,visser2016building,sig10criteria}, which considers only eight guidelines, omitting \emph{i)} and \emph{j)} from above. For each guideline a five star rating is computed by comparing the results within each area with results of analyzing multiple hundreds of undisclosed systems. 
The stars are allocated with a distribution of 5\%-30\%-30\%-30\%-5\%. This means that for a given guideline five stars (best) are assigned when the corresponding measure is among the best 5\% compared to the other systems that \gls{sig_name} uses as reference, four stars when the measure is among the next 30\%, etc. The individual ratings are then aggregated into an overall five star rating for the entire system \cite{visser2016building}. 

\observations{
	It is unclear why the ten guidelines with the corresponding thresholds are chosen. In particular as they seem to be a moving target. During the history of the underlying \gls{sig} (SIG/TÜViT Evaluation Criteria Trusted Product Maintainability) they increased from five source code properties that are measured by the \gls{sig}, see Heitlager et al.~\cite{heitlager2007practical}, over six source code properties, see Baggen et al.~\cite{baggen2012standardized}, to respectively nine in current version of the \gls{sig}~\cite{sig10criteria} and ten such properties in Visser~\cite{visser2016building} and \gls{bch}.
}

\subsection{CAST AIP}\label{subsec:aip}

The company defines \gls{td} as: \emph{``the effort required to fix problems that remain in the code when an application is released.''}\footnote{\url{https://www.castsoftware.com/research-labs/technical-debt-estimation}}. The documentation\footnote{\url{https://doc.castsoftware.com/display/DOC83/Technical+Debt+-+calculation+and+modification}} of \gls{aip} provides a configurable formula for computing \gls{td} in USD.

\begin{equation}
\label{eq:aip_td}
\setlength\abovedisplayskip{0pt}
\scalebox{0.95}[1]{$TD = \left( \Sigma_{i \in \left\{low, medium, high\right\}} r_{i} \times  n_{i} \times  t_{i} \right) \times  c_{staff\_hour}\sfrac{\$}{h}$}
\end{equation}

Where $r$ is the ratio of respectively low, medium, and high severity rules that are violated and that should be fixed. $n$ is the amount of such rule violations, $t$ is the time that it takes to fix a violation of a certain severity. $c_{staff\_hour}$ is the cost for staff to fix issues in USD per hour. In its default configuration, the ratio of low priority violations to fix ($r_{low}$) is set to zero and $c_{staff\_hour}$ is set to 75USD per hour, i.e., the default of \autoref{eq:aip_td} is $TD=((0.5\times n_{medium}\times 0.97h)+(1\times n_{high}\times 2.56h))\times 75\sfrac{\$}{h}$.

To measure \gls{td}, \gls{aip} relies on more than 1\,200 code checking rules in 28 programming languages~\cite{curtis2012estimating}. 
Examples of such rules are \emph{``Avoid Artifacts with High Cyclomatic Complexity`` (Python)} a medium severity rule for keeping \gls{cc} below 20, or the high severity rule \emph{``Avoid header files circular references'' (C)}. 
Severity levels can be adjusted by users.

\gls{aip} defines maintainability independently of \gls{td} as \emph{``the cost and difficulty/ease to maintain an application in the future.''}\footnote{\url{https://doc.castsoftware.com/display/CAST/Glossary\#Glossary-M}} 
and computes it via the \gls{mi}\footnote{\url{https://doc.castsoftware.com/display/TG/CMS+Assessment+Model+-+Information+-+Halstead+metrics+in+CAST+Engineering+Dashboard}}~\cite{foreman1997c4} in two versions:

\begin{align}
\label{eq:aip}
MI_{SEI\_3}&=171-5.2 \times  ln(V_{Hal\_\lambda}) - 0.23 \times  CC_{\lambda} - 16.2 \times  ln(LOC_{\lambda}) \nonumber \\
MI_{SEI\_4}&=MI_{SEI\_3} - 50 \times  sin(\sqrt{2.4 \times  r_{comment\_\lambda}})
\end{align}  

In \autoref{eq:aip}, $V_{Hal\_\lambda}$ is the average Halstead Volume of all modules, which is computed as $V_{Hal} = N \times  ln(\eta)$ per module, where $N$ is the sum of total amounts of operators and operands respectively per module and $\eta$ is the sum of unique operators and operands respectively per module~\cite{foreman1997c4,halstead1975toward}. $CC_{\lambda}$ is the average \gls{cc} (McCabe Complexity~\cite{mccabe1976complexity}) of all modules, which is computed as: $CC=e-n+p$ per module, where $e$ is the number of potentially sequential statements in a module, $n$ is the number of statements in a module, and $p$ is the number of connected components in a graph of sequential statements. $LOC_{\lambda}$ the average lines of code per module. $r_{comment\_\lambda}$ the average ratio of lines of comments per module.

\observations{
	It is unclear if the \gls{mi} are computed as presented in~\cite{foreman1997c4} as that source is only implicitly referenced. Measurement of the values $N,\eta,e,n,p$ for computation of $V_{Hal}$ and $CC$ is unspecified.

	The \gls{td} formula provided in \gls{aip}'s documentation is slightly different from earlier versions reported in~\cite{fontana2016technical,sappidi2010cast,curtis2012estimating_ws,curtis2012estimating}, where it was assumed that all issues take equally long to fix and that the distribution of the amount desired fixes was different. It is unclear why the current values are as they are and whether they will change again.
}

\subsection{Code Climate Quality}\label{subsec:ccq}

\gls{ccq}'s documentation\footnote{\url{https://docs.codeclimate.com/docs}} does not explicitly define the terms maintainability or \gls{td}. But it states that maintainability is the opposite of \gls{td}\footnote{\label{url_climate_td}\url{https://codeclimate.com/blog/10-point-technical-debt-assessment/}} and that \gls{td} \emph{``can be a challenge to measure. Static analysis can examine a codebase for potential structural issues``}. The authors further argue that there \emph{``has never been a single standard, and so we set out to create one.''}\textsuperscript{\ref{url_climate_td}}.

\gls{ccq} computes a \gls{td} ratio as the ratio of total remediation time (also called  \emph{``total technical debt time''}\textsuperscript{\ref{url_climate_td}}, $t_{TD}$) and an estimated time it takes to implement the entire source code ($t_{est\_impl}$):

\begin{equation}
  \text{TD}_{r} = t_{TD} / t_{est\_impl}
\end{equation}

A maintainability rating of each file and the entire system is computed based on $TD_{r}$ as mapping to the discrete values A to F, where A is best, F is worst, and E omitted: A:$\text{TD}_{r}\in\left[0,0.05\right]$, B:$\text{TD}_{r}\in\left]0.05,0.1\right]$, C:$\text{TD}_{r}\in\left]0.1,0.2\right]$, D:$\text{TD}_{r}\in\left]0.2,0.5\right]$, F:$\text{TD}_{r}\in\left]0.5,1\right]$\textsuperscript{\ref{url_climate_td}}. 
$t_{TD}$ is computed based on ten rules, so-called \emph{``\gls{td} checks''} applying to any of the eleven supported programming languages (Ruby, Python, PHP, JavaScript, Java, TypeScript, Go, Swift, Scala, Kotlin, C\#): 
\begin{description}
    \item 1. \emph{argument count} (too many arguments per unit), 
    \item 2. \emph{complex logic} (too long Boolean expressions), 
    \item 3. \emph{file length} (too many lines in a file), 
    \item 4. \emph{identical blocks of code} (syntactic code clones), 
    \item 5. \emph{unit complexity} (units with too high cognitive complexity~\cite{campbell2017cognitive}), 
    \item 6. \emph{unit count} (too many units per modules), 
    \item 7. \emph{method length} (too many lines per unit), 
    \item 8. \emph{nested control flow} (too deeply nested control structures), 
    \item 9. \emph{return statements} (too many return statements per unit), and 
    \item 10. \emph{similar blocks of code} (structural code clones). 
\end{description}
Code can violate any of the ten rules. Each rule has an associated time to fix a violation. The total remediation time $t_{TD}$ is likely just the sum of the remediation time of each violation but that is not explicitly documented. $t_{est\_impl}$ is computed based on the $LOC$. It is unspecified how it is done precisely. We contacted \gls{ccq}'s support for clarification, e.g., on the factor to compute $t_{est\_impl}$ out of $LOC$, but our questions were not addressed.

\observations{
    It is unspecified how the ten \gls{td} checks are precisely computed/implemented for the various languages. Also unspecified is the computation of $t_{est\_impl}$, which remediation cost is associated to each of these rules, if they are the same from language to language, or where they stem from. This makes computation of reported maintainability rating and the \gls{td} ratios/remediation times intransparent.
}

\subsection{Code Inspector}\label{subsec:inspector}

\gls{ci} is likely the youngest tool in this study. We tested its beta version in Nov. 2019. 
\gls{ci} defines \gls{td} via: \emph{``The technical debt Principal is the core issues that might incur rework/interests in the future.''}\footnote{\url{https://www.code-inspector.com/analysis/plan/1961\#collapsePrincipal}}, which it reports as amount of hours and as cost in USD (conversion rate: 70 USD per hour). 
\gls{ci} does neither mention, define nor measure maintainability. 

It is not specified how precisely \gls{td} in hours is computed. We contacted the author and he did not want to reveal the precise computation. 

In the following we describe what we can infer from the online documentation\footnote{\url{http://doc.code-inspector.com/metrics.html}} while analyzing the Java project Apache Commons VFS.
The \gls{td} principal\footnote{\url{https://www.code-inspector.com/analysis/plan/1961}} lists times and cost for remediation of four issue types that are each estimated differently: 
\begin{description}
    \item \emph{Code complexity} as a \gls{cc} value

    \item \emph{Readability} as the length of units

    \item \emph{Duplicated code} as density of number of duplicates per LOC 

    \item \emph{Code violations} as density per LOC of the number of violations of an unspecified amount of rules that are checked on source code\footnote{For each of the eleven supported programming languages (Java, Javascript, C, C++, Go, Ruby, Python, PHP, Scala, Shell scripts, Typescript)}. Examples of such rules are style of variable names, undocumented functions, too many parameters per function, etc. 
\end{description}

\emph{Code violations} are associated to severities \emph{critical}, \emph{major}, \emph{medium}, and \emph{low}.

It is stated that missing documentation on functions has a low impact on \gls{td}, whereas inefficient code or code that might induce buffer overflows has a high impact on \gls{td}, suggesting that different code violations influence \gls{td} differently. However, no more precise computations are specified. Additionally, it is unspecified how the mapping of measurement results to the values \emph{good}, \emph{warning}, and \emph{critical}, via various thresholds\footnote{\url{http://doc.code-inspector.com/metrics.html}} influence the final \gls{td} values.

\observations{
	The actual formulas for computing the \gls{td} (and any other given metric) are neither readily accessible nor documented in a way that allow to understand how analysis of \gls{td} works precisely. The reported \gls{td} values represent the expertise of the tool's authors. 
}

\subsection{CodeScene}\label{subsec:codescene}

\gls{cs} --~one of the youngest tools~-- 
is the only tool in this study that bases its analysis dually on information from \gls{vcs} history and source code artifacts. 
The authors call the inclusion of \gls{vcs} history, \emph{behavioral analysis} and they argue that: \emph{``
we need to consider the temporal dimension of the codebase to avoid spending valuable time improving parts of the code that won't have an impact.''}~\cite{tornhill2018prioritize}.

\gls{cs} defines \gls{td} as a
    \emph{``metaphor that lets developers explain the need for refactorings and communicate technical trade-offs to business people. [\ldots] Just like its financial counterpart, technical debt incurs interest payments.''}~\cite{tornhill2018software}. 
No other definition can be found in the related publications~\cite{tornhill2018assessing,tornhill2018prioritize,tornhill2015your,tornhill2018software}.
\gls{cs} does neither define nor measure maintainability.

Unlike other tools, \gls{cs} does not compute \gls{td} as a single value.
Instead the tool groups the four feedback categories listed below as \gls{td} in its UI:

\begin{enumerate}
    \item \emph{Hotspots} are modules with high complexity (LOC~\footnote{\url{https://codescene.io/docs/guides/technical/hotspots.html}}) and development activity (number of commits), which are visualized using a circle packing layout\footnote{\url{https://codescene.io/projects/174/jobs/17668/results/code/hotspots/system-map}}~\cite{wang2006visualization}. They are considered to serve as a proxy for both \gls{td} and the interest on it~\cite{tornhill2018assessing}.

    \item \emph{Hotspot Code Health}, lists a set of \emph{code biomarkers} for the worst hotspots. Biomarkers are a set of code quality and maintainability metrics --~the author wants to avoid using the terms quality and maintainability\footnote{\url{https://codescene.io/docs/guides/technical/biomarkers.html}}~-- for detecting: code duplication, low cohesion, long methods, deeply nested logic, modularity issues, overall code complexity, etc. Biomarkers are computed on a scale from zero to ten and they are mapped to discrete color codes (red, yellow, green) for the current state of refactoring targets and their states during the last month and year. It is unspecified which code biomarkers exist for which language, how they are computed precisely, and what the corresponding thresholds are that lead to a certain color code.

    \item \emph{Temporal Coupling} (i.e., how often a file was committed with others) is used to indicate hidden architectural coupling in a code base~\cite{tornhill2015your} and is listed under \gls{td} too. It is not indicated how that influences precisely the computation of refactoring targets or how precisely it is related to \gls{td}.

    \item \emph{Refactoring Targets} are computed by an unspecified algorithm from the list of hotspots. In the computation, parameters such as amount of temporally coupled files, amount of affected developers/teams, and likeliness of being a bottleneck for other developers are included.

\end{enumerate}

\observations{
	As refactoring targets and biomarkers are computed intransparently, \gls{cs} merely recommends files worth refactoring instead of assessing \gls{td} of an entire system. 
	The book~\cite{tornhill2015your} gives code examples for some of the performed computations but it remains unclear to which degree these are actually performed by \gls{cs}.

	\gls{cs} seems to rely on three complexity measures at various points of the analysis~\cite{tornhill2018assessing}: 
	    Hotspots seems to be identified via LOC, 
	    complexity trends seem to be based on {\em whitespace complexity}~\cite{hindle2008reading}, 
	    and code biomarkers seem to be based on McCabe complexity\cite{mccabe1976complexity}.
	However, it is not clear nor argued why the various measures are used in these contexts (e.g. they could be also swapped around).

}
\subsection{Kiuwan Code Analysis (QA)}\label{subsec:kiuwan}

\gls{kiuwan} defines maintainability as \emph{``The capability of the software product to be modified.''}\footnote{\url{https://www.kiuwan.com/docs/display/K5/Models+and+CQM}}.
\gls{td} is defined as \emph{``a global effort measure to correct [\ldots] detected defects''}\footnote{\url{https://www.kiuwan.com/docs/display/K5/Governance+Summary}}, where \emph{global} indicates that \gls{kiuwan} can measure \gls{td} for a portfolio of applications and for single applications.

\gls{kiuwan}\footnote{
    \gls{kiuwan} is a commercial product, with which one cannot experiment independently. Instead one has to request a demo with a Kiuwan employee. To avoid a marketing session that could bias our investigation, we experimented with the sibling product ``Kiuwan Code Security (SAST)'', which is accessible for trial, relies on the same quality model CQM, the same rules, and based on investigating the documentation only computes extra security related metrics and ratings. 
} 
measures \gls{td} and maintainability via its own quality model called ``Checking Quality Model for Software'' (CQM), which according to the vendors implements the ISO/IEC 25000 standard focusing on internal quality. Similar to other tools it is based on static source code analysis rules.
\gls{kiuwan} calls rule violations \emph{defects}. For the more than 20 languages that \gls{kiuwan} supports (ranging from ABAP over Cobol, JavaScript, Java, Swift, to VB.NET)\footnote{\url{https://www.kiuwan.com/docs/display/K5/Kiuwan+Supported+Technologies}} per default, more than 1\,300 rules are available, e.g., 216 default rules for Java. Each rule is implemented either as a Java class or via XML\footnote{\url{https://www.kiuwan.com/docs/display/K5/Create+new+Kiuwan+Rules}, \url{https://www.kiuwan.com/docs/display/K5/Getting+Started+with+Rule+Development}}. Associated to rules are one of the five quality characteristics security, reliability, efficiency, maintainability, or portability from the ISO/IEC 25010 standard~\cite{iso25010_2011}. Thus, maintainability is measured as the amount of defects associated with the that quality characteristic\footnote{\url{https://www.kiuwan.com/docs/display/K5/Action+Plans+in+Code+Analysis}}. Out of the 1\,300 default rules 32 are maintainability rules, of which six are for Java.

Additionally, each rule is associated to one of the five severities --~also called priorities~-- \emph{very hard}, \emph{hard}, \emph{normal}, \emph{easy}, and \emph{very easy}, which, per default, are mapped to the effort values 8h, 4h, 30min, 6min, or 3min respectively, indicating the effort in time (man-hours) to fix defects. Without given formulas, \gls{td} said to be the amount of all detected defects weighted by the respective effort in time to fix them\footnote{\url{https://www.kiuwan.com/docs/display/K5/Governance+Summary\#GovernanceSummary-TechnicalDebt}}.

An example for such an \emph{easy to fix} rule is \emph{``Avoid assignments inside conditional expressions''}, which corresponds to CWE-481\footnote{\url{https://cwe.mitre.org/data/definitions/481.html}}. It identifies occurrences of assignments in, e.g., \texttt{if} conditions which can be confused with the more often desired comparison for equality. 

Rules and their parameters, such as effort or priority values, are configurable. Also, they can be created by implementing Java classes or XML files. Next to the default rules, \gls{kiuwan} allows to integrate results of other static analyzers such as PMD, Findbugs, Checkstyle, etc.\footnote{\url{https://www.kiuwan.com/wp-content/uploads/2018/09/Datasheet-Code-Analysis-QA.pdf}}

\observations{
	In \gls{kiuwan} maintainability and \gls{td} are not dialectic concepts. Maintainability measures are aggregated into \gls{td}.
	Even though all rules are listed and described, there is no implementation available for public inspection. 
}
\subsection{NDepend}\label{subsec:ndepend}

NDepend, a tool for C\# and other .Net platform languages\footnote{Although not formally identified by the methodology presented in Section 3.1, the discussion applies also to the sibling products JArchitect for JVM based languages (\url{https://www.jarchitect.com/}) and CppDepend for C++ (\url{https://www.cppdepend.com/}).}, defines \gls{td} as time it takes to fix occurrences of code that do not conform to certain specifications. NDepend does neither define nor measure maintainability.

\gls{td} is measured via 216 static code analysis rules, which are readily accessible\footnote{\url{https://www.ndepend.com/default-rules/NDepend-Rules-Explorer.html}}.
Each rule is expressed in Code Query Linq (CQLinq), i.e., Linq queries that query .Net code via a particular API\footnote{\url{https://www.ndepend.com/docs/cqlinq-syntax}}. Due to space constraints we cannot show the implementation of a rule here, but each rule's implementation is accessible online too. For example, the CQLinq rule \emph{Nested types should not be visible} checks that nested types are declared private\footnote{\label{url_ndepend_rule}\url{https://www.ndepend.com/default-rules/NDepend-Rules-Explorer.html?ruleid=ND1306}}.
When source code violates a rule, NDepend creates a corresponding issue $i$ with associated severity level and \gls{td} $debt(i)$. The latter can be freely computed per rule, see line 12 of the \textit{CQLinq Source Code}\textsuperscript{\ref{url_ndepend_rule}} and the tool's authors define it as \emph{``the estimated man-time that [it] would take to fix the issue''}\footnote{\url{https://www.ndepend.com/docs/technical-debt}}. \gls{td} for the entire source code is measured in man-days and computed as the sum of debt associated with each issue, normalized to man-days ($t_{work} = 8h$ in the formula below being a customizable default):

\begin{equation}
  TD = \left(\Sigma_{i \in Issues} debt(i)\right) / t_{work}
\end{equation}

\gls{td} can be converted to a monetary value where the default price for a man-hour is 50USD, both price and currency are customizable.

NDepend also computes the \emph{annual-interest} ($p_{yr}$) of \gls{td} either via the severities or via a dedicated annual-interest clause per rule. The severity values \emph{info}, \emph{minor}, \emph{major}, \emph{critical}, \emph{blocker} are associated to annual-interest ($p_{yr}$) as: \emph{info}:$p_{yr} \in [0\sfrac{min}{yr},2\sfrac{min}{yr}[$, \emph{minor}:$p_{yr} \in [2\sfrac{min}{yr},20\sfrac{min}{yr}[$, \emph{major}:$p_{yr} \in [20\sfrac{min}{yr},2\sfrac{h}{yr}[$, \emph{critical}:$p_{yr} \in [2\sfrac{h}{yr},10\sfrac{h}{yr}[$, and \emph{blocker}:$p_{yr} \in [10\sfrac{h}{yr},\infty[$.
That is, severities are discrete annual-interest values given in rules. In case continuous values are desired, annual-interest clauses can implement arbitrary computations. The total annual-interest is the sum of the interest declared for all issues. 

NDepends shows the \gls{td} ratio ($TD_{r}$) as a second value for \gls{td} (both the values are listed next to each other under \gls{td}):

\begin{equation}
  TD_{r} = c_{rem} / c_{dev}
\end{equation}

Where $c_{dev}$ is the cost to develop the software and $c_{rem}$ is the remediation cost, i.e., the cost to fix all issues, the $TD$ value computed above. The development cost is computed as $c_{dev} = 8.64min \times LLOC / t_{work}$ as NDepend estimates that it takes 18 man-days to develop 1\,000 logical lines of code (LLOC) that are fully tested and documented.
NDepend operates with the measure of \emph{logical lines of code}, which are inferred out of assembled artifacts\footnote{\url{https://www.ndepend.com/docs/code-metrics\#NbLinesOfCode}}.

A \gls{td} rating is a mapping of the $TD_{r}$ to the discrete values A to E (A is best, E is worst). It is computed as in the following A:$TD_{r} \in [0,0.05[$, B:$TD_{r} \in [0.05,0.1[$, C:$TD_{r} \in [0.1,0.2[$, D:$TD_{r} \in [0.2,0.5[$, and E:$TD_{r} \in [0.5,1]$.
Both \gls{td} ratio $TD_{r}$ and the \gls{td} rating are computed according to the SQALE method~\cite{letouzey2012sqale}.

\observations{
	NDepend offers fine-grained control over computation of \gls{td} via clauses for \gls{td} and annual-interest computation in each rule. 
	Via these rules, the measurement and computation of \gls{td} is transparent.
}

\subsection{SonarQube}\label{subsec:sq}

\gls{sq} defines \gls{td} as the \emph{``Effort to fix all Code Smells.''}\footnote{\label{sq_url_def}\url{https://docs.sonarqube.org/latest/user-guide/metric-definitions/\#header-6}} The effort per ``code smell'' is given in minutes and aggregated across all of the identified smells. When converted to man-days an 8 hour working day is assumed. The term maintainability is not defined explicitly. 

For \gls{sq}, code smells are certain patterns in source code that are considered bad practice, e.g., \emph{``Something that will confuse a maintainer or cause her to stumble in her reading of the code.''}\footnote{\url{https://docs.sonarqube.org/7.8/extend/adding-coding-rules/}}. The default installation of \gls{sq} (version 7.8) contains 1\,740 code smell detection rules for many programming languages (Python 417, Java 340, C\# 261, JavaScript 137, PHP 114, TypeScript 96, VB.NET 89, Flex 64, HTML 41, Kotlin 39, Ruby 38, Scala 38, Go 33, CSS 14, XML 14, JSP 5). For example, the rule \emph{S1871}\footnote{\label{sq_url_rule}\url{https://github.com/SonarSource/sonar-java/blob/62670ebc03aa01346f96d40a2aa999db3487d973/java-checks/src/main/java/org/sonar/java/checks/IdenticalCasesInSwitchCheck.java}} (in the \emph{design} category) checks that 
alternative blocks of \texttt{if} statements should not be the same. Due to space constraints we cannot show the implementation of a rule here, but they are all accessible online\footnote{\url{https://github.com/SonarSource}}. Rules are typically implemented via XPath or Java code via classes containing methods analyzing AST nodes.

Rules are categorized into one of the six remediation effort categories \emph{trivial}, \emph{easy}, \emph{medium}, \emph{major}, \emph{high}, and \emph{complex}, that indicate how hard it is to fix a corresponding issue. The associated remediation effort depends on the programming language that a rule checks. Per default, they are: 5--10min (\emph{trivial}), 10--20min (\emph{easy}), 20--30min (\emph{medium}), one hour (\emph{major}), three hours (\emph{high}), and 8 hours (\emph{complex})\footnote{\url{https://docs.sonarqube.org/7.8/extend/adding-coding-rules/}}. The sum of all remediation efforts, i.e., the total time to fix each detected code smell, forms the \emph{remediation cost} ($c_{rem}$). \gls{sq} calls this cost both \gls{td} and \emph{SQALE index}, using thus three terms for the same concept. Additionally, a \gls{td} ratio ($TD_{r}$) is computed as:

\begin{equation}
TD_{r} = c_{rem} / (c_{per\_line} * LOC)
\end{equation}

where the cost to develop a line of code ($c_{per\_line}$) is set to 0.06 days, with an 8h man-day conversion, $c_{per\_line}\approx30min$.

Using the $TD_{r}$ a \emph{maintainability rating} (also called \emph{SQALE rating}) on a discrete scale from \emph{A} to \emph{E} (\emph{A} is best and \emph{E} is worst) is computed. The documentation states that the mapping from $TD_{r}$ to maintainability rating is: \emph{``A=0-0.05, B=0.06-0.1, C=0.11-0.20, D=0.21-0.5, E=0.51-1''}\textsuperscript{\ref{sq_url_def}}. Consequently, for \gls{sq} \emph{maintainability} is a discretized ratio of \gls{td} per software development cost.

\observations{
	Computing \gls{td} and maintainability rating is done with the SQALE method~\cite{letouzey2016squale}, which provides the formulas above though with different names.
	\gls{sq} provides documentation on how a user can create their own rules. Current rules target source code; but as far as we understand it would be possible to define rules for non-code artifacts.
}
\subsection{SQUORE Software Analytics}\label{subsec:squoring}

For \gls{squore}, \gls{td} is the \emph{``cost of refactoring software to remove all defects and comply with quality requirements''}\cite{squoring} or \emph{``the human effort that shall be invested for the project in order to fix all deviations from the quality standard.''}\footnote{\label{url_demo}\url{https://demo.squore.net/SQuORE_Server/XHTML/MyDashboard/Dashboard.xhtml}}. Maintainability, as in ISO/IEC 14764\cite{iso14764}, is the \emph{``capability of the software to be modified.''}\footnote{\label{url_ref_man}\url{https://demo.squore.net/SQuORE_Server/api/documentation/reference_manual/index.html}}.

\gls{td} is computed based on 1\,857 static code analysis rules --~called the \emph{Squan Sources}~--, which are listed in the manual for 17 programming languages ranging from ABAP to Xaml\textsuperscript{\ref{url_ref_man}}. 
An example for such a rule is \emph{Commented-out Source Code is not allowed}, which raises a \emph{violation} if commented out source code exists. Additionally, \gls{squore} can integrate metrics from 84 tools, such as, PMD, CheckStyle, FindBugs, etc. 

Rules are associated to a \emph{remediation cost} ($c_{rem}$) and so called \emph{ISO characteristics}.
\emph{Remediation costs} can be \emph{tiny}, \emph{low}, \emph{medium}, \emph{high}, and \emph{huge}, which are mapped to time values 1min, 10min, 30min, 1h, and 8h respectively. \emph{ISO characteristics} are the quality characteristics \emph{maintainability}, \emph{reliability}, \emph{efficiency}, \emph{portability}, \emph{security}, \emph{testability}, \emph{changeability}, which --~in this form~-- come from the SQALE model~\cite{letouzey2016squale} which is not an ISO standard.

\gls{squore} computes \gls{td} per artifact, which can be anything from a method, a file, to a package, and entire systems\footnote{\label{url_compu}\url{https://demo.squore.net/SQuORE_Server/api/documentation/swan_handbook/index.html\#sect_computation}}:

\begin{equation}
TD = \sum_{v \in V} (n(v) \times c_{rem}(v))
\end{equation}

where $V$ is the set of all reported rule violations, $n(v)$ the amount of violations of a certain rule and $c_{rem}(v)$ the cost in minutes according to associated remediation cost. \gls{td} is reported either in minutes, hours, or man-days. From inspecting the UI, we observe that 8 hours and 20 minutes correspond to one man-day, which is unspecified in the manuals\footnote{\url{https://demo.squore.net/SQuORE_Server/api/documentation/index.html}}.

Maintainability is computed as the sum of remediation cost of all violations associated to the \emph{maintainability} quality characteristic. The manual is explicit about that \gls{td} and maintainability are two separate concepts\textsuperscript{\ref{url_compu}}. \gls{td} can be computed for the other quality characteristics besides \emph{maintainability}.

\observations{
	\gls{squore} is 
	highly complex with a quite convoluted UI presenting a plethora of measurements
	and the information in the UI and in the manuals is inconsistent. For example, 
	the UI lists the seven quality characteristics \emph{maintainability}, \emph{reliability}, \emph{efficiency}, \emph{portability}, \emph{security}, \emph{testability}, \emph{changeability} where the manual mentions only five omitting \emph{testability} and \emph{changeability}. 
	\gls{td} of some artifacts is reported as, e.g., 7min 30s, where the question arises from where 30s stem from when the lowest remediation cost is set to one minute and the \gls{td} formula operates with integer multiples. 
	It is unspecified how the tools' proprietary rules are implemented. 
}

\subsection{SymfonyInsight}\label{subsec:symfony}

SymfonyInsight is a tool to automate \gls{td} monitoring and quality of PHP based web-applications\footnote{\url{https://insight.symfony.com/}}. 
Maintainability is not explicitly defined by SymfonyInsight and \gls{td} is defined as the \emph{``estimated time a single developer would need to fix all the issues detected by SymfonyInsight''}, which in the tool's UI is also called \emph{remediation cost}. The value for \gls{td} is provided in days, months, or years. It is unspecified how that number is computed and if it is in man-days or plain days. 

Likely, computation is based on the results of applying 110 static analysis rules\footnote{\url{https://insight.symfony.com/what-we-analyse}} (which can be disabled by the user). An example of such a rule is \emph{\#11-013 Database queries should use parameter binding}, which finds code that is prone to SQL injections 
and suggests possible fixes. Rules are organized into seven categories \emph{security}, \emph{architecture}, \emph{performance}, \emph{dead code}, \emph{bug risk}, \emph{readability}, and \emph{coding style}. Each rule has a severity like \emph{info}, \emph{minor}, \emph{major}, and \emph{critical} and a time to fix attached to it. That time does not seem to be related to severity as we find rules in the major category that take 15 minutes vs. two hours to fix. We understand that \gls{td} is just the sum of all the times to fix each single issue. 
The \gls{td} is --~in an unspecified way~-- mapped to five discrete values a quality score given as medals (no medal, bronze, silver, gold, or platinum).

Even though not explicitly defined, a maintainability measure as percentage of change compared to the latest analysis is provided\footnote{\url{https://insight.symfony.com/docs/manager/the-portfolio.html}}. From the UI it appears as if the \emph{dead code} rules correspond to maintainability. \emph{Dead code} comprises of seven rules, including commented out and unreachable code. 
It is unspecified how the change of maintainability is computed.

\observations{
	Using the term remediation cost and mapping quality metrics to discrete measures (medals) suggests that computation of \gls{td} is inspired by the SQALE method. However, without feedback from the vendors, one cannot be sure about this. We contacted the tool vendors, to mitigate high-level documentation and unspecified formulas for \gls{td} and maintainability but we did not receive a reply detailing these.
}

\subsection{Visual Studio}\label{subsec:vs}

According to its documentation\footnote{\url{https://docs.microsoft.com/en-us/visualstudio/?view=vs-2019}}, \gls{vs}
does neither define, measure, nor compute \gls{td}. 
Although the documentation mentions maintainability, the concept is not explicitly defined. However, it is documented to be computed using a modified\footnote{
    \url{https://blogs.msdn.microsoft.com/zainnab/2011/05/26/code-metrics-maintainability-index}} 
version of \gls{mi}, which is calculated as a normalized value between 0 and 100, where higher values represent better maintainability\footnote{
    \url{https://docs.microsoft.com/en-us/visualstudio/code-quality/code-metrics-values?view=vs-2019}}. 
Other than \autoref{eq:aip}, the modified \gls{mi} does not consider comments in source code.

\begin{equation}
\begin{split}
MI = max(0,(171-5.2\times ln(V_{Hal})-0.232\times CC\\
-16.22\times ln(LOC))2\times\sfrac{100}{171})
\end{split}
\end{equation}

\gls{vs} reports a \gls{mi} at various granularities, ranging from project, over file, all the way down to a function or method. When presenting this information, the value is complemented by an icon, which maps maintainability to one of the three ``levels'': high (between 20-100), moderate (between 10-19), and low (between 0-9).

Seemingly simple functions, it is unspecified how exactly $V_{Hal}$, $CC$, and $LOC$ are computed. It is likely that $V_{Hal}$ and $CC$ are implemented as in~\cite{halstead1975toward,mccabe1976complexity} and as described in \autoref{subsec:aip}. Based on the documentation it seems that $CC$ is computed as originally defined: $v=e-n+2p$~\cite{mccabe1976complexity} (which is a difference of factor two compared to the way \gls{aip} computes it, see \autoref{subsec:aip}). 

\observations{
	Transparently, the documentation mentions the use of \gls{mi}. However, the authors stop short of describing how exactly $V_{Hal}$ is implemented. For \gls{cc} the documentation does not specify what \gls{vs} considers operands and operators in C\# and how statement measures in a graph ($e$, $n$, and $p$) are computed precisely. Also, 
	the original \gls{mi} formula~\cite{coleman1994using} was more holistic in that it also considered comments.
}

\begin{table*}
\begin{tabularx}{\textwidth}{>{\hsize=0.01\hsize}X >{\hsize=0.6\hsize}X >{\hsize=0.01\hsize}X >{\hsize=0.01\hsize}X >{\hsize=0.5\hsize}X >{\hsize=0.06\hsize}X >{\hsize=0.06\hsize}X X}
\toprule
ID & Name & \gls{td} & M & Unit & $R_{acc}$ & $R_{imp}$ & Method/Standard \\
\midrule
A & Better Code Hub           & \XSolidBrush & \Checkmark   & Source Code      & \Checkmark   & \XSolidBrush & \gls{sig}, inspired by ISO/IEC 25010 \\
B & CAST AIP                  & \Checkmark   & \Checkmark   & Source Code      & \Checkmark   & \XSolidBrush & proprietary, inspired by ISO/IEC 25010 and ASCMM~\cite{omg2016ascmm} \\
C & Code Climate Quality      & \Checkmark   & \Checkmark   & Source Code      & \Checkmark   & \XSolidBrush & proprietary \\
D & Code Inspector            & \Checkmark   & \XSolidBrush & Source Code      & \XSolidBrush & \XSolidBrush & proprietary \\
E & Codescene                 & \Checkmark   & \XSolidBrush & Source Code, VCS & \XSolidBrush & \XSolidBrush & proprietary \\
F & Kiuwan Code Analysis (QA) & \Checkmark   & \Checkmark   & Source Code      & \Checkmark   & \XSolidBrush & CQM, based on ISO/IEC 25000 \\
G & NDepend                   & \Checkmark   & \XSolidBrush & Source Code      & \Checkmark   & \Checkmark   & SQALE \\
H & SonarQube                 & \Checkmark   & \Checkmark   & Source Code      & \Checkmark   & \Checkmark   & SQALE \\
I & SQUORE Software Analytics & \Checkmark   & \Checkmark   & Source Code      & \Checkmark   & \XSolidBrush & proprietary, based on SQALE and others \\
J & SymfonyInsight            & \Checkmark   & \XSolidBrush & Source Code      & \Checkmark   & \XSolidBrush & proprietary \\
K & Visual Studio             & \XSolidBrush & \Checkmark   & Source Code      & n/a          & n/a & Maintainability Index \\
\bottomrule
\end{tabularx}
\caption{Tools measuring \gls{td} or maintainability (M) on what these are measured (Unit) if all rules are accessible ($R_{acc}$), if rules implementation is accessible $R_{imp}$, and on which method/standard the method is based. (\emph{n/a} for does not apply).}
\label{tab:tools_results}
\end{table*}

\section{Discussion}\label{sec:discussion}

\autoref{tab:tools_results} presents the eleven \gls{td}/maintainability assessment tools. From these, two tools measure solely maintainability (A, K), four tools measure solely \gls{td} (D, E, G, J), and the remaining five tools measure both. That is, more tools provide a measure for \gls{td} than for maintainability.
Six tools (A, %
D, %
E, %
F, %
I, %
J) %
are newly identified in this work compared to previous academic works~\cite{rios2018tertiary,fontana2016technical,lenarduzzi2018survey}. Notably, all eleven tools provide an actual measure of \gls{td} or maintainability

\subsection{RQ1: Defining TD/Maintainability}
Unlike in academia, where there is a trend and effort on settling on a common definition of \gls{td}~\cite{avgeriou2016managing}, there seems to be no such trend in the world of the creators of the studied tools. The tools that provide a measure for \gls{td} provide their own and distinct definitions of it, see \autoref{sec:results}. Even the two tools SonarQube and NDepend that explicitly state that they are based on the same method for managing \gls{td} (SQALE), do not define \gls{td} in the same way. Nevertheless, the \gls{td} definitions of CAST AIP, Kiuwan Code Analysis (QA), NDepend, and SonarQube are conceptually similar in that they consider \gls{td} to be: the cost/time/effort it takes to fix a certain set of problems/issues/defects in code. However, the formulas for computing \gls{td} and estimating times do not share these similarities. Of the seven tools that measure maintainability only three define it explicitly, and each defines it differently as the cost and difficulty to maintain an application (B), the capability of software to be modified (I), and as an estimate of \gls{td} (C).
Consequently, the users of any of the tools must ensure that their understanding of \gls{td} or maintainability is aligned with that of the tool builder. 

\subsection{RQ2: Measuring TD/Maintainability}
All nine tools that measure \gls{td} (except of CodeScene) are based on the concept that a set of static analysis rules --~ranging from 10 to over 1\,000~-- check source code for certain undesired patterns. These rules are either categorized into severities with associated remediation cost (B, H, I, J) or have remediation costs assigned directly on a per-rule basis (C and G). The way the remediation costs are aggregated and converted to other values is different from tool to tool. 

All \gls{td} measures are computed on source code only. Only CodeScene enhances the static code analysis by including a social aspect of software engineering via the analysis of VCS histories. Such a low-level understanding of \gls{td} may differ from the mental model of stakeholders applying a tool. For example, Ernst et al.~\cite{ernst2015measure} demonstrated that software engineers might misinterpret the results presented by the tools since they would expect them to express more architectural issues of the software.

The maintainability measures seem to fall into three categories, either \emph{a)} based on a variation of the \gls{mi}~\cite{oman1992metrics,coleman1994using} (B and K), \emph{b)} based on a set of static analysis rules associated to maintainability (A, F, I), or \emph{c)} as a mapping from \gls{td} measures to discrete values (C and H).

Interestingly, two tools (B and K) still apply the \gls{mi} to assess maintainability. The index is not undisputed~\cite{heitlager2007practical,kuipers2007maintainability,sjoberg2012questioning}, for example, because it \emph{``does not provide clues on what characteristics of maintainability have contributed to that value, nor [\ldots] what action to take to improve this value.''}~\cite{heitlager2007practical}. Additionally, simpler metrics such as LOC or whitespace complexity are proposed as proxies for it~\cite{hindle2008reading}.

Six tools (A, B, F, G, H, I) are either based on SQALE~\cite{letouzey2012managing}, which is based on the ISO/IEC 25000~\cite{iso25000} series of standards, or they are inspired by these two. However, all these tools actually only consider the internal quality model of ISO/IEC 25010. But that standard actually contains another ``quality in use'' model and a ``data quality model'' in ISO/IEC 25012. The latter two models are not regarded by any of the tools, which might lead to a wrong understanding of the reasons of \gls{td} as there seems to be a dualism between program complexity as for example measured by the McCabe Complexity and a corresponding data model.

\subsection{Implications}

None of the tools in this study provides a rationale for why certain static analysis rules are applied or why particular parameters for aggregation are appropriate. 
Moreover, definitions and ways of measuring \gls{td} or maintainability are diverse across vendors and tools, and  customers applying these tools have to trust that given a vendor knows how to identify \gls{td} or maintainability. It would be in the interest of the vendor to be as transparent as possible to increase the confidence of the customer. 

As discussed in \autoref{sec:results}, multiple tools allow the user to configure which rules are active while analyzing software. Some tools (e.g., G and H) allow users even to implement custom rules. With such tools one could --~as advised by SQALE or ISO/IEC 25000~-- \emph{``start by making a list of nonfunctional requirements that define the `right code' [architecture, etc.]''}~\cite{letouzey2012managing}, specify how to assess these requirements, and decide how to aggregate them into higher-level reports or map them to quality characteristics, such as, maintainability. Thereby, one can assure that all stakeholders share a common understanding of \gls{td} or maintainability. Otherwise, when applied directly, i.e., without specifying nonfunctional requirements a-priori, all tools will generate values similar to Tornhill's anecdotal and opaque 4\,000 years.

In fact, it is possible that one of the reasons of the relative popularity of SonarQube compared to other tools in this study\footnote{\url{https://trends.google.com/trends/explore?q=SonarQube,NDepend,CAST\%20AIP,Kiuwan\%20Code\%20Analysis,SQUORE\%20Software\%20Analytics}} is that measurements and computations in it are traceable and configurable. That is, it is transparent how values are measured based on which input data and how precisely they are aggregated into higher level measures. Thereby, users have the possibility to tailor, and be aware about the qualities that contribute to \gls{td} and maintainability. 
To our understanding the only two tools in this study that are configurable and traceable are SonarQube and NDepend.

However management might be inclined to apply one of the studied tools directly to increase maintainability and decrease \gls{td} in the hope that developers that know their work is observed 
will likely think twice before implementing ``not quite right code'' similar to people that change their online behavior when they know they are surveilled~\cite{penney2016chilling}.

\subsection{Threats to Validity} 

Likely, there exist more tools for assessing \gls{td} or maintainability. However, by receiving the input for our study from academic and industrial sources, we believe that we covered a wide-range of potential tools, especially since our list of tools extends previous academic work~\cite{rios2018tertiary,fontana2016technical,lenarduzzi2018survey} substantially. After also inspecting the tools from~\cite{lenarduzzi2018survey}, we believe that we would not have found fundamentally different ways --~only more variation in formulas~-- for measuring and computing \gls{td} or maintainability than those described above.

We might have misunderstood or misinterpreted available documentation of the tools. We tried to carefully gather and understand all available information 

When in doubt we tried to contact the vendors with our questions and examples via their official support channels as indicated in the respective sub-sections of \autoref{sec:results}.

\section{Future Work}\label{sec:conclusions}

Since the contribution of this paper is to properly understand how the identified tools actually assess \gls{td} or maintainability, we have to refer a deeper analysis of similarities and differences across tools as well as a study of possible convergence to a more uniform understanding of \gls{td} to future work.

Furthermore, we plan an experiment in which we apply the listed tools to assess \gls{td} and maintainability of a set of predefined systems. The goal is to quantify how much these measures vary depending on the chosen tool for the same system.

\bibliographystyle{spmpsci}      %
\bibliography{bibliography}   %

\begin{thebibliography}{10}
\providecommand{\url}[1]{{#1}}
\providecommand{\urlprefix}{URL }
\expandafter\ifx\csname urlstyle\endcsname\relax
  \providecommand{\doi}[1]{DOI~\discretionary{}{}{}#1}\else
  \providecommand{\doi}{DOI~\discretionary{}{}{}\begingroup
  \urlstyle{rm}\Url}\fi

\bibitem{sig10criteria}
{SIG/TÜViT Evaluation Criteria Trusted Product Maintainability, Version 12.0}.
\newblock
  \url{https://www.softwareimprovementgroup.com/wp-content/uploads/2020-SIG-TUViT-Evaluation-Criteria-Trusted-Product-Maintainability.pdf}.
\newblock Accessed: 2020-08-01

\bibitem{squoring}
Technical debt squore.
\newblock Tech. rep., Squoring Technologies SAS, FRANCE.
\newblock
  \urlprefix\url{https://www.squoring.com/wp-content/uploads/2014/07/fiche-squoring-technical-debt-en.pdf}

\bibitem{ft2013jfs}
{Aktstykke nr. 127 Folketinget 2012-13}.
\newblock Tech. rep., Folketinget, Copenhagen, DK (2013).
\newblock
  \urlprefix\url{https://www.ft.dk/RIpdf/samling/20121/aktstykke/aktstk127/20121_aktstk_afgjort127.pdf}

\bibitem{ft2015efi}
{Aktstykke nr. 167 Folketinget 2014-15 (2. samling)}.
\newblock Tech. rep., Folketinget, Copenhagen, DK (2015).
\newblock
  \urlprefix\url{https://www.ft.dk/RIPdf/samling/20142/aktstykke/aktstk167/20142_aktstk_afgjort167.pdf}

\bibitem{omg2016ascmm}
Automated source code cisq maintainability measure.
\newblock Standard 1.0, Object Management Group, Inc. (OMG) (2016).
\newblock \urlprefix\url{https://www.omg.org/spec/ASCMM/1.0/PDF}

\bibitem{digst2019model}
{Vejledning til model forporteføljestyring afstatslige it-systemer }.
\newblock Manual, Digitaliseringsstyrelsen, Copenhagen, DK (2019).
\newblock
  \urlprefix\url{https://digst.dk/media/21352/vejledning-til-model-for-portefoeljestyring-af-statslige-it-systemer1-kopi.pdf}

\bibitem{avgeriou2016managing}
Avgeriou, P., Kruchten, P., Ozkaya, I., Seaman, C.: Managing technical debt in
  software engineering (dagstuhl seminar 16162).
\newblock In: Dagstuhl Reports, vol.~6. Schloss Dagstuhl-Leibniz-Zentrum fuer
  Informatik (2016)

\bibitem{baggen2012standardized}
Baggen, R., Correia, J.P., Schill, K., Visser, J.: Standardized code quality
  benchmarking for improving software maintainability.
\newblock Software Quality Journal \textbf{20}(2), 287--307 (2012)

\bibitem{birchall2016re}
Birchall, C.: Re-Engineering Legacy Software.
\newblock Manning Publications Co.

\bibitem{boehm1976quantitative}
Boehm, B.W., Brown, J.R., Lipow, M.: Quantitative evaluation of software
  quality.
\newblock In: Proceedings of the 2nd international conference on Software
  engineering, pp. 592--605. IEEE Computer Society Press (1976)

\bibitem{campbell2017cognitive}
Campbell, G.A.: Cognitive complexity-a new way of measuring understandability.
\newblock Tech. rep., SonarSource SA, Switzerland (2017)

\bibitem{coleman1994using}
Coleman, D., Ash, D., Lowther, B., Oman, P.: Using metrics to evaluate software
  system maintainability.
\newblock Computer \textbf{27}(8), 44--49 (1994)

\bibitem{cunningham1992wycash}
Cunningham, W.: The wycash portfolio management system pp. 29--30 (1992).
\newblock \doi{10.1145/157709.157715}.
\newblock \urlprefix\url{http://doi.acm.org/10.1145/157709.157715}

\bibitem{curtis2012estimating}
Curtis, B., Sappidi, J., Szynkarski, A.: Estimating the principal of an
  application's technical debt.
\newblock IEEE software \textbf{29}(6), 34--42 (2012)

\bibitem{curtis2012estimating_ws}
Curtis, B., Sappidi, J., Szynkarski, A.: Estimating the size, cost, and types
  of technical debt.
\newblock In: Proceedings of the Third International Workshop on Managing
  Technical Debt, pp. 49--53. IEEE Press (2012)

\bibitem{ernst2012role}
Ernst, N.A.: On the role of requirements in understanding and managing
  technical debt.
\newblock In: Proceedings of the Third International Workshop on Managing
  Technical Debt, pp. 61--64. IEEE Press (2012)

\bibitem{ernst2015measure}
Ernst, N.A., Bellomo, S., Ozkaya, I., Nord, R.L., Gorton, I.: Measure it?
  manage it? ignore it? software practitioners and technical debt.
\newblock In: Proceedings of the 2015 10th Joint Meeting on Foundations of
  Software Engineering, ESEC/FSE 2015, pp. 50--60. ACM, New York, NY, USA
  (2015).
\newblock \doi{10.1145/2786805.2786848}.
\newblock \urlprefix\url{http://doi.acm.org/10.1145/2786805.2786848}

\bibitem{fontana2016technical}
Fontana, F.A., Roveda, R., Zanoni, M.: Technical debt indexes provided by
  tools: A preliminary discussion.
\newblock In: 2016 IEEE 8th International Workshop on Managing Technical Debt
  (MTD), pp. 28--31. IEEE (2016)

\bibitem{foreman1997c4}
Foreman, J., Gross, J., Rosenstein, R., Fisher, D., Brune, K., et~al.: C4
  software technology reference guide-a prototype.
\newblock Tech. rep. (1997)

\bibitem{garcia2016improved}
Garc{\'\i}a-Munoz, J., Garc{\'\i}a-Valls, M., Escribano-Barreno, J.: Improved
  metrics handling in sonarqube for software quality monitoring.
\newblock In: Distributed Computing and Artificial Intelligence, 13th
  International Conference, pp. 463--470. Springer (2016)

\bibitem{halstead1975toward}
Halstead, M.H.: Toward a theoretical basis for estimating programming efforts
  (1975)

\bibitem{heitlager2007practical}
Heitlager, I., Kuipers, T., Visser, J.: A practical model for measuring
  maintainability.
\newblock In: 6th international conference on the quality of information and
  communications technology (QUATIC 2007), pp. 30--39. IEEE (2007)

\bibitem{hindle2008reading}
Hindle, A., Godfrey, M.W., Holt, R.C.: Reading beside the lines: Indentation as
  a proxy for complexity metric.
\newblock In: 2008 16th IEEE International Conference on Program Comprehension,
  pp. 133--142. IEEE (2008)

\bibitem{iso25010_2011}
{ISO Central Secretary}: Systems and software engineering -- systems and
  software quality requirements and evaluation (square) -- system and software
  quality models.
\newblock Standard ISO/IEC 25010:2011, International Organization for
  Standardization, Geneva, CH (2011).
\newblock \urlprefix\url{https://www.iso.org/standard/35733.html}

\bibitem{iso14764}
{Software Engineering -- Software Life Cycle Processes -- Maintenance}.
\newblock Standard, International Organization for Standardization, Geneva, CH
  (2006).
\newblock
  \urlprefix\url{https://www.iso.org/obp/ui/#iso:std:iso-iec:14764:ed-2:v1:en}

\bibitem{iso25000}
{Systems and software engineering -- Systems and software Quality Requirements
  and Evaluation (SQuaRE) -- Guide to SQuaRE}.
\newblock Standard, International Organization for Standardization, Geneva, CH
  (2014).
\newblock \urlprefix\url{https://www.iso.org/obp/ui/\#!iso:std:64764:en}

\bibitem{kaiser2011selling}
Kaiser, M., Royse, G.: Selling the investment to pay down technical debt: The
  code christmas tree.
\newblock In: 2011 Agile Conference, pp. 175--180. IEEE (2011)

\bibitem{kitchenham1996software}
Kitchenham, B., Pfleeger, S.L.: Software quality: the elusive target [special
  issues section].
\newblock IEEE software \textbf{13}(1), 12--21 (1996)

\bibitem{kuipers2007maintainability}
Kuipers, T., Visser, J.: Maintainability index revisited--position paper.
\newblock In: Special session on system quality and maintainability (SQM 2007)
  of the 11th European conference on software maintenance and reengineering
  (CSMR 2007). Citeseer (2007)

\bibitem{lenarduzzi2018survey}
Lenarduzzi, V., Sillitti, A., Taibi, D.: A survey on code analysis tools for
  software maintenance prediction.
\newblock In: International Conference in Software Engineering for Defence
  Applications, pp. 165--175. Springer (2018)

\bibitem{letouzey2012sqale}
Letouzey, J.L.: The sqale method for evaluating technical debt.
\newblock In: 2012 Third International Workshop on Managing Technical Debt
  (MTD), pp. 31--36. IEEE (2012)

\bibitem{letouzey2016squale}
Letouzey, J.L.: The sqale method for managing technical debt definition
  document  (2016).
\newblock
  \urlprefix\url{http://www.sqale.org/wp-content/uploads/2016/08/SQALE-Method-EN-V1-1.pdf}

\bibitem{letouzey2012managing}
Letouzey, J.L., Ilkiewicz, M.: Managing technical debt with the sqale method.
\newblock IEEE software \textbf{29}(6), 44--51 (2012)

\bibitem{mccabe1976complexity}
McCabe, T.J.: A complexity measure.
\newblock IEEE Transactions on software Engineering (4), 308--320 (1976)

\bibitem{oman1992metrics}
Oman, P., Hagemeister, J.: Metrics for assessing a software system's
  maintainability.
\newblock In: Proceedings Conference on Software Maintenance 1992, pp.
  337--344. IEEE (1992)

\bibitem{penney2016chilling}
Penney, J.W.: Chilling effects: Online surveillance and wikipedia use.
\newblock Berkeley Tech. LJ \textbf{31}, 117 (2016)

\bibitem{rios2018tertiary}
Rios, N., de~Mendonca~Neto, M.G., Sp{\'\i}nola, R.O.: A tertiary study on
  technical debt: Types, management strategies, research trends, and base
  information for practitioners.
\newblock Information and Software Technology \textbf{102}, 117--145 (2018)

\bibitem{dos2013visualizing}
dos Santos, P.S.M., Varella, A., Dantas, C.R., Borges, D.B.: Visualizing and
  managing technical debt in agile development: An experience report.
\newblock In: International Conference on Agile Software Development, pp.
  121--134. Springer (2013)

\bibitem{sappidi2010cast}
Sappidi, J., Curtis, B., Subramanyam, J.: Cast worldwide application software
  quality study — 2010.
\newblock Tech. rep., CAST Software Inc., New York (2010).
\newblock
  \urlprefix\url{https://www.agilealliance.org/wp-content/uploads/2016/01/CAST_2010AnnualReport_KeyFindings_WebFinal.pdf}

\bibitem{sjoberg2012questioning}
Sj{\o}berg, D.I., Anda, B., Mockus, A.: Questioning software maintenance
  metrics: a comparative case study.
\newblock In: Proceedings of the 2012 ACM-IEEE International Symposium on
  Empirical Software Engineering and Measurement, pp. 107--110. IEEE (2012)

\bibitem{tornhill2015your}
Tornhill, A.: Your code as a crime scene: use forensic techniques to arrest
  defects, bottlenecks, and bad design in your programs.
\newblock Pragmatic Bookshelf (2015)

\bibitem{tornhill2018assessing}
Tornhill, A.: Assessing technical debt in automated tests with codescene.
\newblock In: 2018 IEEE International Conference on Software Testing,
  Verification and Validation Workshops (ICSTW), pp. 122--125. IEEE (2018)

\bibitem{tornhill2018prioritize}
Tornhill, A.: Prioritize technical debt in large-scale systems using codescene.
\newblock In: Proceedings of the 2018 International Conference on Technical
  Debt, TechDebt '18, pp. 59--60. ACM, New York, NY, USA (2018).
\newblock \doi{10.1145/3194164.3194187}.
\newblock \urlprefix\url{http://doi.acm.org/10.1145/3194164.3194187}

\bibitem{tornhill2018software}
Tornhill, A., Tulton, A.: Software Design X-Rays: Fix Technical Debt with
  Behavioral Code Analysis.
\newblock Pragmatic programmers. Pragmatic Bookshelf (2018)

\bibitem{visser2016building}
Visser, J., Rigal, S., van~der Leek, R., van Eck, P., Wijnholds, G.: Building
  Maintainable Software, Java Edition: Ten Guidelines for Future-Proof Code,
  1st edn.
\newblock O'Reilly Media, Inc. (2016)

\bibitem{wang2006visualization}
Wang, W., Wang, H., Dai, G., Wang, H.: Visualization of large hierarchical data
  by circle packing.
\newblock In: Proceedings of the SIGCHI Conference on Human Factors in
  Computing Systems, CHI '06, pp. 517--520. ACM, New York, NY, USA (2006).
\newblock \doi{10.1145/1124772.1124851}.
\newblock \urlprefix\url{http://doi.acm.org/10.1145/1124772.1124851}

\end{thebibliography}

\end{document}